\documentclass{aa} 
\usepackage{amsmath}
\usepackage{graphicx}
\usepackage{txfonts}
\usepackage{color}
\begin{document}
   \title{Hint of a truncated primordial spectrum from the CMB large-scale anomalies}

   \author{F. Melia$^1$\thanks{John Woodruff Simpson Fellow}, Qingbo Ma$^2$, Jun-Jie Wei$^3$ and Bo Yu$^4$} 

   \offprints{F. Melia}
\titlerunning{The CMB large-scale anomalies}
\authorrunning{Melia, Ma, Wei \& Yu}

\institute{$^1$Department of Physics, The Applied Math Program, and Department of Astronomy, 
The University of Arizona, Tucson, Arizona 85721, USA; \email{fmelia@email.arizona.edu} \\
$^2$Guizhou Provincial Key Laboratory of Radio Astronomy and Data Processing, and
School of Physics and Electronic Science, Guizhou Normal University, Guiyang 550001, China \\
$^3$PMO, Nanjing 210023, China; Guangxi Key Laboratory for Relativistic Astrophysics, Guangxi 
University, Nanning 530004, China; University of Chinese Academy of Sciences, Beijing 100049, China \\
$^4$PMO, Nanjing 210023, China}

   \date{Received May 4, 2021}

 
  \abstract
   {Several satellite missions have uncovered a series of potential anomalies in the fluctuation
spectrum of the cosmic microwave background temperature, including: (1) an unexpectedly low
level of correlation at large angles, manifested via the angular correlation function,
$C(\theta)$; and (2) missing power in the low multipole moments of the angular power
spectrum, $C_\ell$.}
   {Their origin is still debated, however, due to a persistent lack of
clarity concerning the seeding of quantum fluctuations in the early Universe. A
likely explanation for the first of these appears to be a cutoff,
$k_{\rm min}=$~$(3.14\pm0.36)\times 10^{-4}$ Mpc$^{-1}$, in the primordial power
spectrum, ${\mathcal{P}}(k)$. Our goal in this paper is twofold: (1) we examine whether the
\hbox{\it same} $k_{\rm min}$ can also self-consistently explain the missing power at large angles,
and (2) we confirm that the introduction of this cutoff in ${\mathcal{P}}(k)$ does not
adversely affect the remarkable consistency between the prediction of \hbox{\it Planck}-$\Lambda$CDM
and the \hbox{\it Planck} measurements at $\ell>30$.}
   {We use the publicly available code CAMB to calculate the angular power spectrum, based on a 
line-of-sight approach. The code is modified slightly to include the additional parameter 
(i.e., $k_{\rm min}$) characterizing the primordial power spectrum. In addition to this cutoff, 
the code optimizes all of the usual standard-model parameters.}
   {In fitting the angular power spectrum, we
find an optimized cutoff, $k_{\rm min}=(2.04^{+1.4}_{-0.79})\times 10^{-4}$~Mpc$^{-1}$, when
using the whole range of $\ell$'s, and $k_{\rm min}=(3.3^{+1.7}_{-1.3})\times 10^{-4}$~Mpc$^{-1}$,
when fitting only the range $\ell\le 30$, where the Sachs-Wolfe effect is dominant.}
   {These are
fully consistent with the value inferred from $C(\theta)$, suggesting that both of these
large-angle anomalies may be due to the same truncation in ${\mathcal{P}}(k)$.}


   \maketitle
%

\section{Introduction} 
All three major satellite missions designed to study the cosmic microwave background
(CMB)---COBE (Hinshaw et al. 1996), WMAP (Bennett et al. 2003), and {\it Planck}
(Planck Collaboration VII 2020)---have uncovered several unexpected features in its
temperature fluctuations. These include missing correlations at large angles and
an unexpectedly low power in multipoles $2\lesssim \ell\lesssim 30$, especially
$\ell=2$ and $3$. In spite of these apparent deficiencies, however, the agreement
between the angular power spectrum predicted by $\Lambda$CDM and the observations
is quite remarkable for $\ell>30$, forming the basis for the optimization of
many cosmological parameters.

The large-scale anomalies stand in sharp contrast to the general level of success
interpreting the CMB anisotropies, sustaining a simmering debate concerning their origin,
or possible misidentification due to unknown systematic effects (see Bennett et al. 2011,
Copi et al. 2010 for reviews). If these two unexpected features are real,
it is not even clear if they are due to the same physical process, in spite of the fact
that they both characterize variations in temperature over very large angular scales.

A thorough analysis of these anomalies was carried out with the final
release of the {\it Planck} data by the Planck Collaboration VII (2020). They studied
the statistical isotropy and Gaussianity of the CMB using {\it both} the Planck 2018 
temperature and polarization data. Previous work had focused solely on the temperature
fluctuations, due to only a limited ability to also probe the CMB polarization. 
The large-angular-scale polarization measurements permit a largely independent 
examination of the peculiar features seen in the temperature which can, in principle, 
reduce or eliminate any subjectivity or ambiguity in interpreting their statistical
significance.

The possibility that the lack of large-angle correlations in the temperature may be largely
due to a statistical `fluke,' as unlikely as that may appear to be, was raised by the fact 
that the Planck Collaboration VII study found only weaker evidence of an analogous lack of 
large-scale angular correlations in the polarization data.

Nevertheless, in spite of the fact that no unambiguous detections of cosmological 
non-Gaussianity, or of anomalies analogous to those seen in temperature, could be claimed,
efforts at reducing the systematic effects that contaminated the earlier polarization
maps on large angular scales could not completely\break\vfill\newpage\noindent exclude the 
presence of residual systematics that limit some tests of non-Gaussianity and isotropy. 

A comparison of both the temperature and polarization analyses thus tends to mitigate our
ability to conclude one way or the other whether the large-angle anomalies are real. From
a theoretical standpoint, the polarization fluctuations also originated in the primordial 
gravitational potential, but are largely sourced by different modes than the temperature
anisotropies. Thus, a measurement of large-scale anomalies in {\it both} the temperature
and polarization has the potential of providing a greater statistical significance. It is
important to see whether any anomalies seen in the polarization maps are related to known
features in the temperature. On the other hand, if no such anomalies are seen in the
polarization it might be the case that the temperature anomalies, if not pure flukes,
could be due to secondary influences, such as the integrated Sachs-Wolfe effect
(Planck Collaboration XXI 2016), or more exotic scenarios in which they may be due 
to physical processes that do not correlate with the polarization.

The situation with regard to the large-angle anomalies is thus far from clear, and
it is essential to continue broadening the study of possible causes of such features
in either the temperature, polarization or both. In earlier work (Melia \& 
L\'opez-Corredoira 2018), we showed that a likely theoretical explanation for the 
missing large-angle correlations in temperature is the presence of a cutoff, 
$k_{\rm min}$, in the  primordial power spectrum, ${\mathcal{P}}(k)$. Indeed,
our analysis of the {\it Planck} data revealed that a zero cutoff is ruled out by
these measurements at a confidence level $\gtrsim 8\sigma$. Instead, we found that
the observed angular correlation function, $C(\theta)$, could be reproduced
satisfactorily at all angles with a cutoff $k_{\rm min}=(4.34\pm0.50)/r_{\rm dec}$,
where $r_{\rm dec}$ is the comoving distance between us and the last scattering
surface at decoupling ($z_{\rm dec}=1080$). A truncation such as this is not
necessarily consistent with the aims of slow-roll inflation to simultaneously
fix the horizon problem and seed a quantum fluctuation spectrum consistent with
large-scale structure (Liu \& Melia 2020, Melia 2020), because $k_{\rm min}$ represents the
time at which inflation could have started, severely constraining the number of
e-folds available for the Universe to expand at an accelerated rate. It is therefore
imperative to pursue this line of inquiry further and see if additional evidence
may be found in favor of a non-zero $k_{\rm min}$.

The goal of this paper is to complete that analysis in two significant ways: (1)
to examine whether the same cutoff may be responsible for both large-angle anomalies,
and (2) to confirm that the introduction of $k_{\rm min}$ in ${\mathcal{P}}(k)$
does not adversely impact the remarkable agreement between theory and observations
at $\ell>30$. In \S~2 we introduce the basic theoretical background underlying this
work, and then proceed to study the effects of a truncated ${\mathcal{P}}(k)$ on
the angular power spectrum in \S~3. We discuss the results in \S~4, and present
our conclusion in \S~5.

\section{Theoretical background}
The inflationary paradigm posits that an early accelerated expansion persisted long
enough to create a homogeneous and isotropic Universe within our current Hubble volume.
As such, the CMB temperature, $T$, measured in some direction $\hat{e}$, is expected
to be a Gaussian random field on the sky, expandable as a series of spherical harmonics,
$Y_{\ell m}(\hat{e})$, with independent Gaussian random coefficients $a_{\ell m}$ of zero mean:
\begin{equation}
\langle a_{\ell m}\rangle=0\;,
\end{equation}
for all $\ell>0$ and $m=-\ell,-\ell+1,...,+\ell$.
If the Universe is indeed statistically isotropic, this spherical harmonic decomposition,
and the two-point angular power spectrum derived from it, contain all of the physical
information needed to interpret the CMB anisotropies.

The less used (but equally important) angular correlation function $C(\theta)$ relates the
temperature measured in two independent directions, $\hat{e}_1$ and $\hat{e}_2$, and depends
solely on the dot product $\cos\theta\equiv\hat{e}_1\cdot\hat{e}_2$. One may thus expand it
in terms of Legendre polynomials:
\begin{equation}
\langle T(\hat{e}_1)T(\hat{e}_2)\rangle\equiv C(\theta)={1\over 4\pi}\sum_\ell (2\ell+1)\,C_\ell
P_\ell(\cos\theta)\;,\label{eq:angcorr}
\end{equation}
in which the variance $C_\ell$ is known as the angular power of multipole $\ell$. Statistical
independence implies that the expectation of a product of $a_{\ell m}$'s with different
$\ell$'s and $m$'s vanishes, while isotropy requires the constant of proportionality for this
product to be solely a function of $\ell$, not $m$:
\begin{equation}
\langle a_{\ell m}^*\,a_{{\ell}^{\prime} {m}^{\prime}}\rangle=
\delta_{\ell {\ell}^{\prime}}\,\delta_{m m^\prime}\,C_\ell\;.
\end{equation}

\begin{figure*}[h]
\vskip 1cm
\centerline{
\includegraphics[angle=0,scale=0.6]{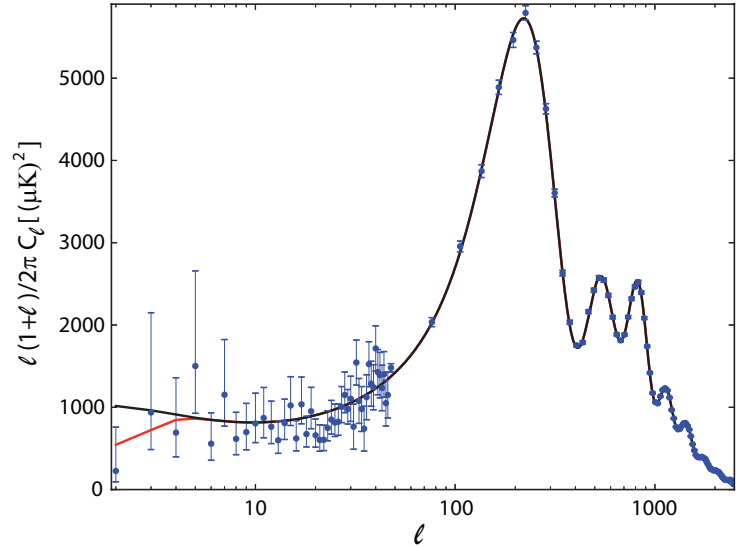}}
\vskip 0.2in
\caption{Angular power spectrum (blue dots) estimated with the {\it NLIC} method, with $1\sigma$ 
Fisher errors. These error bars include cosmic variance, approximated as Gaussian 
(see fig.~1 in Planck Collaboration VI 2020). The $\Lambda$CDM best fit model 
($C_\ell^{TT}$), with a power-law fluctuation distribution and $k_{\rm min}=0$, is shown in black. 
The red curve (for $k_{\rm min}=2.82/r_{\rm dec}$) is the best fit to the whole range of $\ell$'s 
with an optimized $k_{\rm min}\not=0$. The two curves are indistinguishable at $\ell\gtrsim 6$, 
but deviate significantly from each other for the low multipole moments, corresponding to large 
angles ($\gtrsim 60^\circ$). The Sachs-Wolfe contribution to $C_\ell^{TT}$ dominates at 
$\ell\lesssim 30$, while local physical effects, such as acoustic oscillations, dominate 
at $\ell\gtrsim 30$.}
\label{f1}
\end{figure*}

As one can see from figure~\ref{f1} (discussed below), the standard model has been strikingly successful
in accounting for the angular power spectrum (i.e., $C_\ell$ versus $\ell$), at least on
small angular scales ($\lesssim 1^\circ$), corresponding to $\ell\gtrsim 100$. The rather
precise determination of the various cosmological parameters (Planck Collaboration VI 2020) is
based on this excellent agreement between theory and observation.

In principle, the angular correlation function $C(\theta)$ contains the same information as the
angular power spectrum (see Eq.~\ref{eq:angcorr}), but there are good reasons for analyzing the CMB
anisotropies separately using these two approaches. First, whereas the angular power spectrum
highlights the relative contributions of different spherical harmonics, the angular correlation
function describes the variations of $T$ in real space. Some features may emerge more prominently
in one description instead of the other.  Second, $C(\theta)$ highlights properties of the
anisotropies at relatively large angles (i.e., small $\ell$'s), while the power spectrum provides
ample detail in the opposite regime, namely $\ell\gg 10$. 
One can easily see the reason for this with a quick inspection of figure~\ref{f1}. When using
the full angular power spectrum, the model fit is dominated by over 1,000 values of $\ell$ at
$\ell>30$, and while the optimized fit is striking in this regime, as we discuss elsewhere in 
this paper, the error bars and variance at $\ell<30$ are much larger and the fit is significantly
less compelling. On the other hand, the angular correlation function displays the temperature
anisotropies with equal weighting over all angles $0\le\theta\le 180^\circ$. When fitting a
model to these data, the dominant influence is therefore at $\theta>10^\circ$, the opposite
of what happens with the power spectrum. Third, $C(\theta)$ is a
direct pixel-based measure, not requiring any reconstruction of the observed sky. Some of the
systematics may therefore have less impact in one approach relative to the other.

In spite of these differences, one may nonetheless reasonably expect that, of the various
large-angle anomalies seen in the CMB (Planck Collaboration VII 2020), the one 
associated with $C(\theta)$ could have the same
physical origin as a second emerging from the angular power spectrum. Many authors (see, e.g.,
Copi et al. 2010) have pointed out that the most striking feature of the angular
correlation function (as seen, say, in fig.~2 of Melia \& L\'opez-Corredoira 2018) is not
just that it disagrees with $\Lambda$CDM at a very high level of confidence ($\gtrsim 3\sigma$;
Copi et al. 2015), but that it is nearly zero at large angles ($\gtrsim 60^\circ$)---in significant
tension with inflation, which should have produced correlations throughout the visible Universe.

Most of the statistical weight in the {\it Planck} analysis of the angular power spectrum
is provided by the high-$\ell$ multipoles, and one may see in figure~\ref{f1}
that the best-fitting portion of the curve traces the data at $\ell\gtrsim 60$ much better
than those at low-$\ell$'s. Indeed, a general lack of angular power appears within the
range $2\lesssim \ell\lesssim 30$---especially with the quadrupole $\ell=2$---at a confidence
level exceeding $99\%$ (Bennett et al. 2011, Planck Collaboration VI 2020). Quite remarkably,
none of the attempted fixes invoking physically motivated inputs, such as neutrino properties,
the number of relativistic degrees of freedom, or a running of the spectral index in the
fluctuation distribution, have provided any significant improvement to the fit (see,
e.g., Planck Collaboration XXIV 2014; Planck Collaboration VI 2020). To be sure, the $C(\theta)$ 
anomaly is much more obvious than the low harmonic-space quadrupole and octopole power (see also 
O'Dwyer et al. 2004) but, given that they both reference CMB anisotropic structure at very large 
angles ($\gtrsim 60^\circ$), it is tempting to consider the possibility that these two anomalies 
may have a common physical origin (see discussion in Planck Collaboration VII 2020).

This is still not generally accepted, however. The low power at large angles may simply be
due to cosmic variance (Efstathiou 2003b), in which our Universe may represent a random
extrapolation away from its most likely configuration. Another possible explanation for the
vanishing $C(\theta)$ at large angles is that a {\it range} of low multipoles are `conspiring'
to cause the cancellation (Copi et al. 2009). In this interpretation, the combined contribution
of $C_2$--$C_5$ is canceled by the contributions of $C_\ell$ with $\ell>5$, an apparent
conspiracy that seems to violate the independence of different multipole powers expected
in statistical isotropy. But all such attempts at reconciling the missing angular correlations
with low power at $\ell\lesssim 30$ have thus far been inconclusive. Sentiment has therefore
shifted to the idea that finding a theoretical explanation for the missing power at low
$\ell$'s may be best served by first explaining the vanishing $C(\theta)$ at large angles.

\begin{figure*}[h]
\vskip 1cm
\centerline{
\includegraphics[angle=0,scale=0.6]{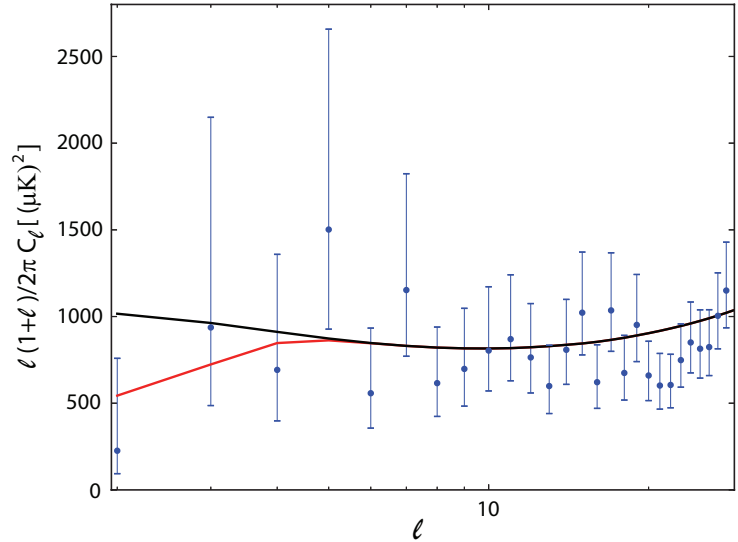}}
\vskip 0.2in
\caption{Same as figure~\ref{f1}, except magnifying the view at $\ell\lesssim 30$, where the dominant
contribution to $C_\ell^{TT}$ is the Sachs-Wolfe effect. The best-fit curves are indistinguishable
for $\ell\gtrsim 6$, but the comparison with the data is greatly improved for the low-$\ell$
multipoles.}
\label{f2}
\end{figure*}

\section{Best-fit truncated spectrum}
\subsection{The angular correlation function}
We recently addressed this issue (Melia \& L\'opez-Corredoira 2018) by attempting to
show that the absence of large-angle correlations in the {\it Planck} data is best explained
by the introduction of a non-zero minimum wavenumber $k_{\rm min}$ in the primordial
fluctuation power spectrum,
\begin{equation}
{\mathcal{P}}(k) = \begin{cases} A_s\left({k\over k_0}\right)^{n_s-1} & \mbox{if } k\ge k_{\rm min}  \\
0 & \mbox{if } k<k_{\rm min} \end{cases}\;,\label{eq:power}
\end{equation}
where $n_s$ is the spectral index, measured to have the value $0.965\pm0.004$ by {\it Planck}
(Planck Collaboration VI 2020); $A_s=2.196^{+0.051}_{-0.06}\times 10^{-9}$ is the spectral
amplitude; and $k_0=0.05$ Mpc$^{-1}$ is the scalar pivot scale. Such a cutoff challenges
inflation's ability to simultaneously fix the horizon problem and produce a near scale-free
distribution of quantum fluctuations matching the observed CMB anisotropies (Liu \& Melia 2020; 
Melia 2020), because the presence of a non-zero $k_{\rm min}$ implies a specific time at which 
the accelerated expansion could have started, thus restricting the number of e-folds available 
to expand the Universe beyond the Hubble horizon. In contrast, standard inflationary 
$\Lambda$CDM has no cutoff---and hence no limit (in principle)---to the degree of
inflated expansion. This is the reason, of course, why inflation predicts an angular
correlation at all scales which, however, has not been observed by any of the CMB
satellite missions (Wright et al. 1996; Bennett et al. 2003; Planck Collaboration VI 2020).

Our earlier analysis of the angular correlation function, based on the notion that quantum
fluctuations were generated in the early Universe with this power spectrum and
$k_{\rm min}\not=0$, demonstrated that the {\it Planck} data rule out a zero cutoff
at a confidence level exceeding $8\sigma$. An optimized fit to $C(\theta)$
showed that the correlations measured over the {\it entire} sky could be reproduced
remarkably well with a cutoff
\begin{equation}
k_{\rm min}={4.34\pm0.50\over r_{\rm dec}}\;,
\end{equation}
where $r_{\rm dec}$ is the comoving distance between us and redshift $z_{\rm dec}=1080$,
at which decoupling is thought to have occurred in standard $\Lambda$CDM. Using the latest
{\it Planck} parameters (Planck Collaboration VI 2020), one finds $r_{\rm dec}\approx 13,804$~Mpc,
and therefore a corresponding minimum wavenumber
\begin{equation}
k_{\rm min}=(3.14\pm 0.36)\times 10^{-4}\;{\rm Mpc}^{-1}\;\;\;({\rm ang.\; corr.\; 
function})\;,\label{eq:kmin}
\end{equation}
roughly 160 times smaller than the pivot scale $k_0$.

This previous study, however, did not complete the analysis by investigating how such a cutoff
would impact the angular power spectrum---both at low and high $\ell$'s. Given how well
inflationary $\Lambda$CDM reproduces the observed $C_\ell$'s at $\ell\gtrsim 60$ (see fig.~\ref{f1}),
the question in not merely whether a non-zero $k_{\rm min}$ could also reduce the power of the
low multipoles, but whether it could do so while simultaneously preserving the excellent
agreement seen between theory and observation elsewhere in this plot. Answering this question
is the principal goal of this paper.

\subsection{The angular power spectrum}
The idea of reducing power at large angles by modifying the potential of inflation is not
new (for a comprehensive review, see Martin et al. 2013). Other approaches have included:
(1) altering inflation's initial conditions (Berera et al. 1998; Contaldi et al. 2003;
Boyanovsky et al. 2006; Powell \& Kinney 2007; Wang \& Ng 2008; Cicoli et al. 2014;
Das et al. 2015; Broy et al. 2015; Liu \& Melia 2020; Melia 2020); (2) a reconsideration of the Integrated
Sachs-Wolfe effect (Das \& Souradeep 2014); (3) the introduction of spatial curvature
(Efstathiou 2003a); (4) the use of a non-trivial topology (Luminet et al. 2003); (5) 
geometric effects (Campanelli et al. 2006; Campanelli et al. 2007); (6) a violation 
of statistical anisotropies (Hajian \& Souradeep 2003); the creation of primordial micro 
black-hole remnants (Scardigli et al. 2011); and loop quantum cosmology (Barrau et al. 2014), 
among several others. Two particular approaches stand out in terms of their overlap with 
our treatment in this paper, notably those of Iqbal et al. (2015) and Santos et al. (2018), 
to which we shall return in \S~4 below, when we discuss our results in a broader context.

The analysis we carry out here is novel for two principal reasons: (i)
it is complementary to---and required for the completion of---our previous study of the
angular correlation function (Melia \& L\'opez-Corredoira 2018), and (ii) it is very
focused on the specific modification to the primordial fluctuation spectrum exhibited
in Equation~(\ref{eq:power}), which is itself highly motivated by its success in resolving the
missing correlations at large angles. Can a truncated spectrum ${\mathcal{P}}(k)$ with
the same $k_{\rm min}$ resolve both the $C(\theta)$ and low-$\ell$ anomalies?

\begin{figure*}[t]
\vskip 1cm
\centerline{
\includegraphics[angle=0,scale=0.6]{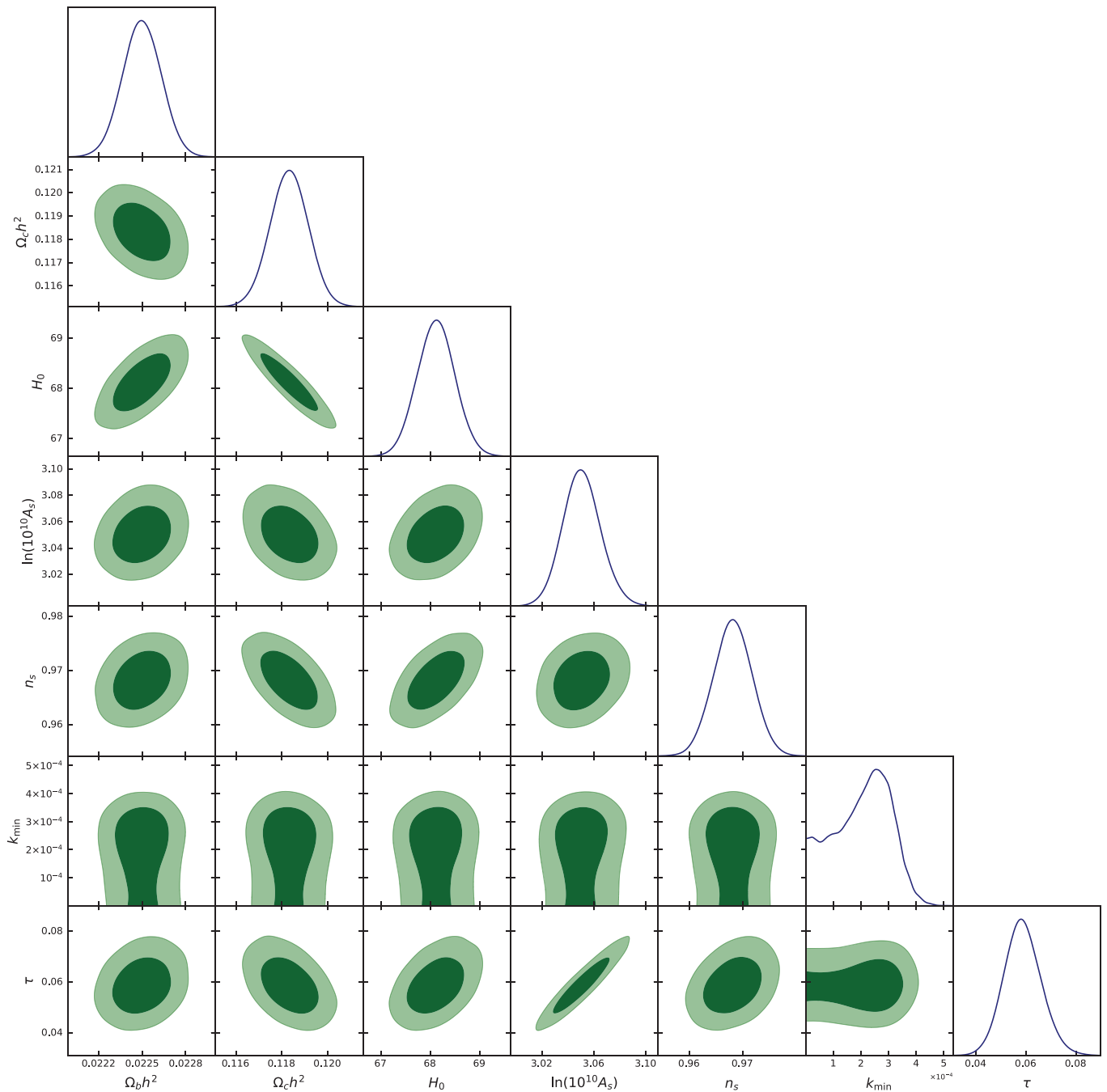}}
\vskip 0.2in
\caption{The six basic parameters in $\Lambda$CDM optimized in our procedure,
based on the combined data sets SN+DES+BAO+{\it Planck}, together with
the new parameter ($k_{\rm min}$) we have introduced to represent a cutoff in the
primordial power spectrum, ${\mathcal{P}}(k)$.}
\label{f3}
\end{figure*}

We use the publicly available code CAMB\footnote{www.camb.info} (Lewis et al. 2000) to
calculate the angular power spectrum, based on a line-of-sight approach described in
Seljak \& Zaldarriaga (1996). The code is modified slightly to include the additional
parameter (i.e., $k_{\rm min}$) characterizing the primordial power spectrum (Eq.~\ref{eq:power}).
In addition to this cutoff, the code optimizes all of the usual standard-model
parameters but, as we shall see from an inspection of figures~\ref{f1}--\ref{f3} and Table~\ref{table1},
most of them remain unchanged from their {\it Planck}-$\Lambda$CDM values. We therefore
highlight here a comparison only of the six basic parameters between the two cases in which
$k_{\rm min}=0$ and $k_{\rm min}\not=0$: baryon density, $\Omega_{\rm b}$, cold
dark matter density, $\Omega_{\rm c}$, the Thomson scattering optical depth, $\tau$,
due to reionization, the angular size of the acoustic horizon, $\theta_s$, the
spectral index, $n_s$, and the scalar amplitude, $A_s$. Each of the quantities
$\Omega_i$ represents the ratio of the energy density $\rho_i$ of species ``i"
to the critical density, $\rho_{\rm c}\equiv 3c^2 H_0^2/8\pi G$, in terms of the
Hubble constant $H_0$. Our optimization procedure is even less sensitive to
the other parameters in {\it Planck}-$\Lambda$CDM.

In optimizing the fit, we use the publicly available code cosmomc (Lewis \& Bridle 2002)
with an initial cutoff range $0\le k_{\rm min}\le 10^{-3}$ Mpc$^{-1}$, and a combined data 
set that includes: the low-$\ell$ TT and EE and high-$\ell$ TT, TE and EE likelihoods from 
Planck 2018 (Planck Collaboration VI 2020), the Dark Energy Survey (DES) Year 1 results
(Abbott et al. 2018; Troxel et al. 2018), the Baryon Acoustic Oscillation (BAO) compilation
from Alam et al. (2017), and the Pantheon Type Ia SN catalog (Scolnic et al. 2018). 
Note that the error bars shown in figures~\ref{f1} and \ref{f2} include cosmic variance,
approximated as Gaussian (see fig.~1 in Planck Collaboration VI 2020), and so do the 
likelihoods also taken from Planck Collaboration VI (2020) used with our cosmomc calculations.

The calculated angular power $C_\ell$ for {\it Planck}-$\Lambda$CDM, with
$k_{\rm min}=0$, is indicated by the black curve in figures~\ref{f1} and \ref{f2}, in
comparison with the {\it Planck} data (blue dots). Our corresponding optimized fit using
$k_{\rm min}$ as a free parameter is represented by the red solid curve.
Based on this fit, we find an optimized value
\begin{equation}
k_{\rm min}={2.82^{+1.94}_{-1.09}\over r_{\rm dec}}\;\;\;({\rm ang.\; power\; spectrum})\;,
\end{equation}
corresponding to a maximum fluctuation size $\theta_{\rm max}\approx 127^\circ$
in the plane of the sky. For a redshift $z_{\rm dec}=1080$ at decoupling, this
cutoff represents a maximum fluctuation size $\lambda_{\rm max}\sim 31$ Mpc at
that redshift. The six basic parameters for this fit, and their confidence regions,
are shown in figure~\ref{f3} and listed in Table~\ref{table1}. Note that,
in spite of the fact that a $k_{\rm min}$ compatible with zero appears to be possible 
within $1\sigma$ in the $\tau-k_{\rm min}$ panel of figure~\ref{f3}, this is not
the case because the $1\sigma$ and $2\sigma$ regions are different in 1D and 2D.
The distinction arises because the $1\sigma$ and $2\sigma$ levels in the 2D histograms 
are not the $68\%$ and $95\%$ values used for the 1D distributions.  The relevant $1\sigma$
and $2\sigma$ levels for a 2D histogram of samples are $1-\exp(-0.5)\rightarrow 39.3\%$ and 
$1-\exp(-2)\rightarrow 86.5\%$.\footnote{https://corner.readthedocs.io/en/latest/pages/sigmas.html.} 
Using $39.3\%$ and $86.5\%$ of the region, the contour plots then do exclude a $k_{\rm min}$ 
compatible with zero at the $1\sigma$ (1D) level.

\begin{figure*}[t]
\vskip 1cm
\centerline{
\includegraphics[angle=0,scale=0.6]{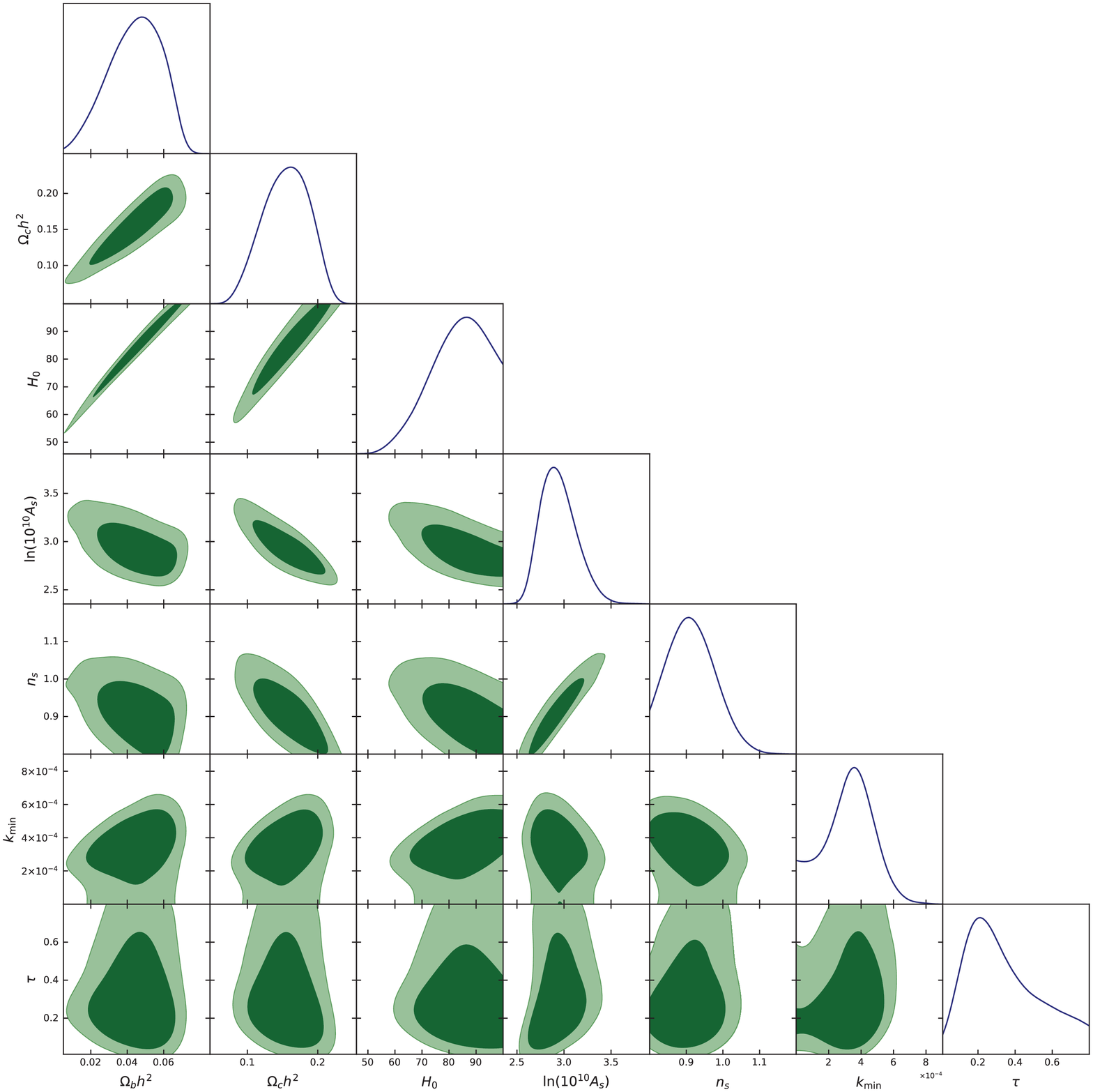}}
\vskip 0.2in
\caption{Same as figure~\ref{f3}, except now using an optimization of the fit for
$\ell\le 30$, and the low-$\ell$ TT data from {\it Planck} 2018 (Planck Collaboration VI 2020),
combined with the DES, BAO, and Type Ia SNe data sets.}
\label{f4}
\end{figure*}

The fact that a non-zero cutoff $k_{\rm min}$ has minimal, if any, impact 
on the optimized fit at $\ell\gtrsim 30$, is consistent with the prevailing view that
the physical origin of the fluctuations is due to different physical effects
above and below $\sim 1^\circ$. It may therefore be useful to compare this
optimized value of $k_{\rm min}$ based on the whole range of $\ell$'s with that found
when $\ell$ is restricted to the range ($\lesssim 30$) thought to be dominated by the
Sachs-Wolfe effect (Sachs \& Wolfe 1967). The optimized parameter values for this case, using
the low-$\ell$ TT data from {\it Planck} 2018, combined with the DES, BAO, and Type Ia
SNe data sets, are shown in figure~\ref{f4}. The corresponding best-fit cutoff $k_{\rm min}$
for this restricted range is $(3.3^{+1.7}_{-1.3})\times 10^{-4}$ Mpc$^{-1}$, consistent
with the full-$\ell$ range value, but even closer to the cutoff found using the angular
correlation function. In either case, the cutoff `fixing' the low-power anomaly at
$\ell\lesssim 4$ is fully consistent with the value that completely accounts for the
missing correlations at angles $\gtrsim 60^\circ$, suggesting that both of these
large-angle anomalies in the standard model are due to the same truncation in
${\mathcal{P}}(k)$.

\begin{table}
\centering \caption{Optimized Parameters in $\Lambda$CDM with and without $k_{\rm min}\not=0$}
\begin{tabular}{lcc}
\hline
\hline
&&\\
Parameters &  $k_{\rm min}=0$  &  $k_{\rm min}\not=0$  \\
&&\\
\hline
&&\\
$H_0$ (${\rm km}\;{\rm s}^{-1}\;{\rm Mpc}^{-1}$)&  $68.12\pm 0.37$ & $68.12\pm0.38$\\
$\Omega_{\rm b}\,h^2$ & $0.02250\pm 0.00013$ & $0.02250\pm 0.00013$\\
$\Omega_{\rm c}\,h^2$ & $0.11830\pm 0.00082$ & $0.11832\pm 0.00082$\\
$\tau$ & $0.0586^{+0.0067}_{-0.0078}$ &  $0.0587 ^{+0.0069}_{-0.0077}$\\
$n_{\rm s}$ & $ 0.9682\pm 0.0036$ & $0.9681 \pm 0.0036$\\
$\ln(10^{10}A_{\rm s})$ & $3.050^{+0.013}_{-0.015}$ & $3.050 \pm 0.015$\\
$k_{\rm min}$ (${\rm Mpc}^{-1}$)& 0 & $(2.04^{+1.4}_{-0.79})\times 10^{-4}$ \\
&&\\
\hline
\end{tabular}
\label{table1}
\end{table}

\section{Discussion}
A side-by-side comparison of the best-fit parameters in $\Lambda$CDM with and without
a cutoff $k_{\rm min}$ (Table~\ref{table1}) reveals only negligible differences between
them, but the angular power spectrum for $\ell\lesssim 4$ is improved significantly
compared with the data. Thus, the excellent agreement between standard $\Lambda$CDM
and the CMB data for $\ell\gtrsim 60$ is completely unaffected by
our introduction of a cutoff to the primordial power spectrum. Yet, based on
the optimized $k_{\rm min}$ value using the whole $\ell$ range (and even more
so using the restricted range $\ell\lesssim 30$), we conclude
that a zero cutoff is ruled out at a confidence level of $\sim 2.6\sigma$.
On its own, this is already interesting enough for us to study its impact
on slow-roll inflation (Liu \& Melia 2020; Melia 2020), but when combined with our earlier
measurement of $k_{\rm min}$ based on the angular correlation function
(Melia \& L\'opez-Corredoira 2018), which showed that $k_{\rm min}=0$ was ruled out at
a confidence level exceeding $8\sigma$ for those data, these two complementary
indicators argue compellingly that the primordial power spectrum
${\mathcal{P}}(k)$ must have been truncated at $k_{\rm min}\sim 3\times 10^{-4}
\;{\rm Mpc}^{-1}$.

These results are completely consistent with the prevailing view that the CMB
anisotropies are primarily due to two distinct physical influences: the Sachs-Wolfe
effect (Sachs \& Wolfe 1967) at small $\ell$'s and acoustic oscillations for $\ell\gtrsim 30$.
The excellent fit to the data for large $\ell$'s (see fig.~\ref{f1}) reinforces the perception
that we understand fairly well the physical process associated with small-angle anisotropies,
in spite of the fact that the angular correlation function has been problematic since the
beginning (Hinshaw et al. 1996; Bennett et al. 2003; Planck Collaboration VI 2020).

The emerging evidence for a non-zero $k_{\rm min}$, on the other hand, speaks directly
to the cosmological expansion itself. At $\ell\lesssim 30$, we see the effects due to
the metric perturbations associated with the cosmological dynamics, and therefore the
direct influence of the cosmological model itself. In this range, we are probing ever
closer to the onset of the hypothesized inflationary expansion, culminating with the cutoff
$k_{\rm min}$, that signals the very first mode crossing the horizon at the beginning
of the quasi-de Sitter phase produced by the inflaton field (Liu \& Melia 2020; Melia 2020).

The idea that at least one of the large-angle anomalies may be due
to features in the primordial power spectrum, ${\mathcal{P}}(k)$, has been with us
for several decades (see discussion in Planck Collaboration VII 2020). For example, 
Shafieloo \& Souradeep (2004) assumed an exponential
cutoff at low $k$'s, and found an optimized fit of the attenuation generally
consistent with the cutoff we have measured in this paper. Their ansatz for
${\mathcal{P}}(k)$ did not include an actual truncation, however, which would be
required for self-consistency with our previous treatment of $k_{\rm min}$ in
Melia \& L\'opez-Corredoira (2018). The impact on inflation is different in these two cases,
since a sharp termination signals a precise time at which inflation would have
started. An exponential cutoff is not fully consistent with the notion of
horizon-crossing for the freezing of modes. In addition, this early treatment
was based on the use of WMAP data (Bennett et al. 2003), whereas our 
analysis incorporates the higher precision {\it Planck} observations (Planck 
Collaboration VI 2020).  In figure~4, we have used only the {\it Planck} TT data 
at $\ell\le 30$, plus other low-redshift data sets, as described in the caption.
For figure~3, however, we used all of the {\it Planck} data, including TT, TE, EE 
and also both the high-$\ell$ and other low-redshift data sets. A comparison of these
results shows that the optimized value of $k_{\rm min}$ is very similar in these
two cases, suggesting that the other low-redshift data sets, such as DES, BAO and 
Type Ia SNe, do not significantly affect the limits on the cutoff.

In related work, Nicholson \& Contaldi (2009) and Hazra et al. (2014) modeled a `dip'
in ${\mathcal{P}}(k)$ on a scale $k\sim 0.002$ Mpc$^{-1}$ for the WMAP data.
Their results confirmed those of an alternative approach (Ichiki et al. 2010),
in which an oscillatory modulation was identified around $k\sim 0.009$ Mpc$^{-1}$.
These are not a truncation, however, nor are they consistent with our analysis
(Melia \& L\'opez-Corredoira 2018) of a cutoff $k_{\rm min}$ in the angular correlation function
based on the {\it Planck} measurements.

Tocchini-Valentini et al. (2005) modeled both a dip at $k\sim 0.035$ Mpc$^{-1}$ and a `bump'
at $k\sim 0.05$ Mpc$^{-1}$, also using WMAP data. As with the others, however,
these features are not the same as an actual truncation $k_{\rm min}$, and our
analysis utilizes the latest {\it Planck} data, rather than the less precise WMAP measurements.
Subsequent work by these authors, notably Tocchini-Valentini et al. (2006), improved on
this analysis considerably, though they were still fully reliant on WMAP.
These authors found evidence for three features in
${\mathcal{P}}(k)$, one of which is analogous to our $k_{\rm min}$,
including a cutoff at $\sim 0.0001-0.001$ Mpc$^{-1}$, with a confidence
level of about $2\sigma$. Our analysis based on the improved {\it Planck}
data produces a tighter constraint on this cutoff, more in line with
the $k_{\rm min}$ found from the angular correlation function. For example,
their possible cutoff range exceeds the actual value we have measured for
$k_{\rm min}$ in this paper.

More recently, Hunt \& Sarkar (2014, 2015) found evidence in
the WMAP data of a cutoff $k < 5 \times 10^{-4}$ Mpc$^{-1}$ to
${\mathcal{P}}(k)$, but concluded that more accurate data, such as those
provided by Planck, would need to be used to confirm these results more
robustly. The work most closely aligned with our analysis in this paper
is that of Iqbal et al. (2015) and Santos et al. (2018). These two groups
arrived at opposite conclusions to each other, however. Though each carried out
a thorough analysis of the possible cause for the low CMB power at small $\ell$'s,
and actually agreed on approximately what value a cutoff (which they called
$k_c$, analogous to our $k_{\rm min}$) would be needed to address this anomaly,
they adopted different model-selection criteria and produced opposing outcomes.

This state of uncertainty has been with us for many years. The approach of
Iqbal et al. (2015) and Santos et al. (2018) was to focus solely on the low-$\ell$
anomaly and to compare 7 or 8 different models, using a range of model-selection
tools. Our approach in this paper has been diametrically opposite to this. We have
only one goal in mind, motivated by the physics of inflation. Most importantly,
our previous analysis demonstrated that a zero value for $k_{\rm min}$ is ruled
out at over $8\sigma$. Notice, e.g., how different this outcome is compared to
that of Santos et al. (2018), who concluded that standard inflationary cosmology,
without any adjustment to ${\mathcal{P}}(k)$, is preferred by the CMB data and
their choice of model selection criteria.

The work we have carried out in this paper is based on the conclusion of our previous 
work suggesting that only a simple cutoff $k_{\rm min}$ in ${\mathcal{P}}(k)$ is well 
motivated by the CMB angular correlation function. We have not compared different 
models, nor have we based our analysis on a subjective choice of model selection 
criteria, which seem to produce the divergent outcomes seen, e.g., in Iqbal et al. (2015) 
and Santos et al. (2018).

We showed in Liu \& Melia (2020) that the cutoff $k_{\rm min}$ measured via the CMB angular
correlation function apparently rules out all inflationary models based on slow-roll
potentials, including those that may have been preceded by radiation-dominated and/or
kinetic-dominated phases (see also Handley et al. 2014). With such a cutoff, 
inflation could not simultaneously have
solved the horizon problem and accounted for the observed CMB fluctuation spectrum.
This is the reason why a confirmation of $k_{\rm min}$ based on the angular power
spectrum is so essential. The two large-angle anomalies are not necessarily related
(see also Planck Collaboration VII 2020), 
so our previous result with $k_{\rm min}$ does not by itself ensure that it is also consistent
with the much better studied angular power spectrum, nor that it can also mitigate
the missing power at low-$\ell$'s. This paper has therefore been highly focused on
these two issues alone. Nevertheless, the fact that our measured value of $k_{\rm min}$ 
is consistent with that found by Iqbal et al. (2015) and Santos et al. (2018) is crucial 
because it demonstrates a consistent outcome by three different groups.

We mention in passing that an extension of the work reported by Iqbal et al. (2015)
and Santos et al. (2018) to include the angular correlation function would be very
beneficial. Our previous assessment was that only a simple cutoff in ${\mathcal{P}}(k)$
could completely account for the missing correlations at large angles. It would be
interesting to see if an independent examination of these data, using a range of
other models with various model selection tools, such as those explored by these
two groups, could alter this conclusion.

In this paper, we have met our primary goal of showing that the introduction of a
non-zero cutoff in ${\mathcal{P}}(k)$ does not at all affect the excellent agreement
between the predictions of $\Lambda$CDM and the CMB angular power spectrum at $\ell\gtrsim 30$.
This is critical because it is widely believed that the origin of the CMB anisotropies
in this domain is well understood, given that it relies on well-established astrophysical
principles.

We have also demonstrated that the origin of the missing power in the low-$\ell$
multipoles, particularly at $\ell=2$--$5$, is very likely due to the same physical
influence impacting the missing correlations at large angles. The two optimized
values of $k_{\rm min}$, i.e., $(3.14\pm0.36)\times 10^{-4}$ Mpc$^{-1}$ for the latter, and
$(2.04^{+1.4}_{-0.79})\times 10^{-4}$ Mpc$^{-1}$ for the former, are fully
consistent with each other. These complementary results reinforce the view that we are
beginning to see direct evidence of the initiation of inflation, if this process actually
did occur. Nevertheless, the disparity between the primordial fluctuation spectrum expected
under these conditions and what is actually required to produce the CMB anisotropies,
increases the tension between the concordance model and the observations.

\section{Conclusion}
The existence of large-angle anomalies in the temperature is uncontested though, given the
modest significance with which they disagree with the standard model, and the fact that
they were detected a posteriori, makes it unclear how much evidence they actually provide
for a true, physical origin (Planck Collaboration VII 2020). They may simply be statistical 
fluctuations, even though such an outcome appears to be highly unlikely. Nevertheless, if 
any of them do indeed correspond to a real physical effect, it would be extremely important 
for us to confirm this, justifying the continued attention paid to these unusual features.

As noted earlier, however, we must still temper our conclusions regarding the
reality of these large-angle anomalies, given that the Planck Collaboration VII (2020) study 
found only weaker evidence for them in the polarization data. This is why upcoming missions
designed specifically to measure the CMB polarization will be so critical to the continuation
of this work. For example, LiteBIRD has been selected for development and launch in 2028 by
the Japan Aerospace Exploration Agency, with the goal of mapping the CMB polarization over 
the whole sky with unprecedented precision (Hazumi et al. 2019). Its required angular coverage 
will correspond to $2\le\ell\le 200$, perfectly attuned to the requirements for the analysis 
in this paper. 

Perhaps even more impressively, the Probe of Inflation and Cosmic Origins (PICO) mission
(Hanany et al. 2019) is still in the study phase, but is projected to be an imaging polarimeter 
scanning the sky for 5 years in 21 frequency bands spread between 21 and 799 GHz. If selected,
PICO will produce a full-sky survey of the intensity and polarization of the CMB with a final 
combined-map noise level equivalent to an amazing 3,300 Planck missions.

And as a prominent third example, the Cosmic Origins Explorer (CORE) mission (Delabrouille et al.
2018) proposed to the European Space Agency is projected to have 19 frequency channels spanning 
the $60-600$ GHz range, and an angular resolution from $2^\prime$ to $18^\prime$. It will observe 
the entire sky repeatedly over four years of continuous scanning. Its design will be optimized
for complementarity with ground-based observations, performing observations---such as that
required by the analysis carried out in this paper---essential to CMB polarization science 
not achievable without a dedicated space mission.

Proposed, or in-progress, experiments to map the CMB polarization from the ground include the
Simons Observatory (Ade et al. 2019), located in the high Atacama Desert in Northern Chile inside 
the Chajnator Science Preserve, at an altitude of 5,200 meters. Though its goals are currently 
not as lofty as those of the space-based missions, its principal aim is to produce a polarization 
map of the sky with an order of magnitude better sensitivity than {\it Planck}.

The principal aim of all these future missions is, of course, to detect the B-mode
polarization in the tell-tale signature of gravity waves generated during inflation. 
Nevertheless, a more precise measurement of the E-mode polarization anisotropies at a much 
higher sensitivity than that available to {\it Planck} should answer the question of whether
or not the large-angle anomalies seen in the temperature are also confirmed in the polarization
maps. This would then provide compelling evidence that they are not merely due to secondary, 
background or instrumental effects, further motivating a study of the impact on inflationary 
theory of a cutoff, $k_{\rm min}$, in the primordial power spectrum, ${\mathcal{P}}(k)$. 

For example, as shown by Liu \& Melia (2020), a $k_{\min}$ like that measured earlier in our 
analysis of the angular correlation function (Melia \& L\'opez-Corredoira 2018), and reinforced 
by the work reported in this paper, which is based exclusively on the angular power spectrum,
rules out the majority---if not all---of the slow-roll inflaton potentials proposed thus
far. If one insists on inflation simultaneously fixing the horizon problem and accounting
for the observed primordial power spectrum, ${\mathcal{P}}(k)$, then the accelerated
expansion resulting from these hypothesized fields misses the required comoving distance
by a factor $\sim 10$. Moreover, neither a radiation-dominated, nor a kinetic-dominated,
phase preceding inflation can alleviate this disparity (Liu \& Melia 2020; see also
Handley et al. 2014).

Our results reinforce the growing view that, at a minimum, inflation probably needs
to be modified or, at worst, needs to be replaced, in order to conform with these
observations.

\begin{acknowledgements}
We are grateful to the anonymous referee for their helpful and thoughtful review
of this manuscript. FM is also grateful to Amherst College for its support through a John Woodruff
Simpson Lectureship. This work is partially supported by the National Natural Science Foundation
of China (grant No. U1831122), the Youth Innovation Promotion Association (2017366), and the Key
Research Program of Frontier Sciences (grant No. ZDBS-LY-7014) of Chinese Academy of Sciences,
at Purple Mountain Observatory, and by the Innovation and Entrepreneurial Project of
Guizhou Province for High-level Overseas Talents (grant No. 2019-02), the National
Natural Science Foundation of China (grant No. 11903010), and the Science and Technology
Fund of Guizhou Province (grant No. 2020-1Y020), at Guizhou University.
\end{acknowledgements}


\begin{thebibliography}{}
\bibitem{Abbott:2018} {Abbott}, T.M.C., {Abdalla}, F. B., {Alarcon}, A. {et~al.}, 2018, PRD, 98, 043526
\bibitem{Ade:2019} Ade, P. et al., 2019, JCAP, 2019, id. 056
\bibitem{Alam:2017} {Alam}, S., {Ata}, M., {Bailey}, S. {et~al.}, 2017, MNRAS, 470, 2617
\bibitem{Barrau:2014} {Barrau}, A., {Cailleteau}, T., {Grain}, J. \& {Mielczarek}, J., 2014, CQG, 31, 053001
\bibitem{Bennett:2003} {Bennett}, C. L., {Hill}, R. S., {Hinshaw}, G. {et~al.}, 2003, ApJ Sup, 148, 97
\bibitem{Bennett:2011} {Bennett}, C. L., {Hill}, R. S., {Hinshaw}, G. {et~al.}, 2011, ApJ Sup, 192, 17
\bibitem{Berera:1998} {Berera}, A., {Fang}, L. Z. \& {Hinshaw}, G., 1998, PRD, 57, 2207
\bibitem{Boyanovsky:2006} {Boyanovsky}, D., {de Vega}, H. J. \& {Sanchez}, N. G., 2006, PRD, 74, 123006
\bibitem{Broy:2015} {Broy}, B. J., {Roest}, D. \& {Westphal}, A., 2015, PRD, 91, 023514
\bibitem{Campanelli:2006} {Campanelli}, L., {Cea}, P. \& {Tedesco}, L., 2006, PRL, 97, 131302
\bibitem{Campanelli:2007} {Campanelli}, L., {Cea}, P. \& {Tedesco}, L., 2007, PRD, 76, 063007
\bibitem{Cicoli:2014} {Cicoli}, M., {Downes}, S., {Dutta}, B., {Pedro}, F. G. \& {Westphal}, A., 2014,
JCAP, 2014, 030
\bibitem{Contaldi:2003} {Contaldi}, C. R., {Peloso}, M., {Kofman}, L. \& {Linde}, A., 2003, JCAP, 2003, 002
\bibitem{Copi:2009} {Copi}, C. J., {Huterer}, D., {Schwarz}, D. J. \& {Starkman}, G. D., 2009, MNRAS, 399, 295
\bibitem{Copi:2010} {Copi}, C. J., {Huterer}, D., {Schwarz}, D. J. \& {Starkman}, G. D., 2010,
arXiv e-prints, arXiv:1004.5602
\bibitem{Copi:2015} {Copi}, C. J., {Huterer}, D., {Schwarz}, D. J. \& {Starkman}, G. D., 2015,
MNRAS, 451, 2978
\bibitem{Das:2014} {Das}, S. \& {Souradeep}, T., 2014, JCAP, 2014, 002
\bibitem{Das:2015} {Das}, S., {Goswami}, G., {Prasad}, J. \& {Rangarajan}, R., 2015, JCAP, 2015, 001
\bibitem{Delabrouille:2018} Delabrouille J. et al., 2018, JCAP, 2018, id. 014
\bibitem{Efstathiou:2003a} {Efstathiou}, G., 2003a, MNRAS, 343, L95
\bibitem{Efstathiou:2003b} {Efstathiou}, G., 2003b, MNRAS, 346, L26
\bibitem{Hajian:2003} {Hajian}, A. \& {Souradeep}, T., 2003, ApJ Lett, 597, L5
\bibitem{Hanany:2019} Hanany, S. et al., eprint arXiv:1902.10541
\bibitem{Handley:2014} Handley, W. J., Brechet, S. D., Lasenby, A. N., \& Hobson, M. P.,
2014, PRD, 89, id. 063505
\bibitem{Hazra:2014} {Hazra}, D. K., {Shafieloo}, A. \& {Souradeep}, T., 2014, JCAP, 2014, 011
\bibitem{Hazumi:2019} Hazumi, M. et al., 2019, J. Low Temp. Phys., 194, 443
\bibitem{Hinshaw:1996} {Hinshaw}, G., {Branday}, A. J. {Bennett}, C. L. {et~al.}, 1996, ApJ Lett, 464, L25
\bibitem{Hunt:2014} {Hunt}, P. \& {Sarkar}, S., 2014, JCAP, 2014, 025
\bibitem{Hunt:2015} {Hunt}, P. \& {Sarkar}, S., 2015, JCAP, 2015, 052
\bibitem{Ichiki:2010} {Ichiki}, K., {Nagata}, R. \& {Yokoyama}, J., 2010, PRD, 81, 083010
\bibitem{Iqbal:2015} {Iqbal}, A., {Prasad}, J., {Souradeep}, T. \& {Malik}, M, A., 2015, JCAP, 2015, 014
\bibitem{Lewis:2000} {Lewis}, A., {Challinor}, A. \& {Lasenby}, A., 2000, ApJ, 538, 473
\bibitem{Lewis:2002} {Lewis}, A. \& {Bridle}, S., 2002, PRD, 66, 103511
\bibitem{Liu:2020} {Liu}, J. \& {Melia}, F., 2020, Proc R Soc A, 476, 20200364
\bibitem{Luminet:2003} {Luminet}, J. P., {Weeks}, J. R., {Riazuelo}, A., {Lehoucq}, R. \& {Uzan}, J P., 2003,
Nature, 425, 593
\bibitem{Martin:2013} {Martin}, J., {Ringeval}, C. \& {Vennin}, V., 2013, JCAP, 2013, 021
\bibitem{Melia:2018} {Melia}, F., 2018, EPJ-C, 78, 739
\bibitem{Melia:2020} {Melia}, F., 2020, {\em The Cosmic Spacetime} (Taylor \& Francis, Oxford)
\bibitem{MeliaLopez:2018} {Melia}, F. \& {L{\'o}pez-Corredoira}, M., 2018, A\&A, 610, A87
\bibitem{Nicholson:2009} {Nicholson}, G. \& {Contaldi}, C. R., 2009, JCAP, 2009, 011
\bibitem{O'Dwyer:2004} {O'Dwyer}, I. J., {Eriksen}, H. K., {Wandelt}, B. D. {et~al.}, 2004, ApJ Lett, 617, L99
\bibitem{Planck:2014} Planck Collaboration XXIV, Ade, P.A.R. et~al., 2014, A\&A, 571, A24
\bibitem{Planck:2016} {Planck Collaboration XXI}, Ade, P.A.R., 2016, A\&A, 594, A21
\bibitem{PlanckVI:2020} {Planck Collaboration VI}, {Aghanim}, N., {Akrami}, Y. {et~al.}, 2020, A\&A, 641, A6
\bibitem{PlanckVII:2020} {Planck Collaboration VII}, Akrami, Y. et al., 2020, A\&A, 641, A7
\bibitem{Powell:2007} {Powell}, B. A. \& {Kinney}, W. H., 2007, PRD, 76, 063512
\bibitem{Sachs:1967} {Sachs}, R. K. \& {Wolfe}, A. M., 1967, ApJ, 147, 73
\bibitem{Santos:2018} {Santos da Costa}, S., {Benetti}, M. \& {Alcaniz}, J., 2018, JCAP, 2018, 004
\bibitem{Scardigli:2011} {Scardigli}, F., {Gruber}, C. \& {Chen}, P., 2011, PRD, 83, 063507
\bibitem{Scolnic:2018} {Scolnic}, D. M., {Jones}, D. O., {Rest}, A. {et~al.}, 2018, ApJ, 859, 101
\bibitem{Seljak:1996} {Seljak}, U. \& {Zaldarriaga}, M., 1996, ApJ, 469, 437
\bibitem{Shafieloo:2004} {Shafieloo}, A. \& {Souradeep}, T., 2004, PRD, 70, 043523
\bibitem{Tocchini:2005} {Tocchini-Valentini}, D., {Douspis}, M. \& {Silk}, J., 2005, MNRAS, 359, 31
\bibitem{Tocchini:2006} {Tocchini-Valentini}, D., {Hoffman}, Y. \& {Silk}, J., 2006, MNRAS, 367, 1095
\bibitem{Troxel:2018} {Troxel}, M. A., {MacCrann}, N., {Zuntz}, J. {et~al.}, 2018, PRD, 98, 043528
\bibitem{Wang:2008} {Wang}, I. C. \& {Ng}, K. W., 2008, PRD, 77, 083501
\bibitem{Wright:1996} {Wright}, E. L., {Bennett}, C. L., {Gorski}, K., {Hinshaw}, G. \& {Smoot}, G. F., 1996,
ApJ Lett, 464, L21
\end{thebibliography}
\end{document}